\documentclass[a4paper,11pt]{article}
\usepackage{amsmath,amssymb,amsthm}
\usepackage[english]{babel}

\usepackage[pdftex]{graphicx}\def\grext{pdf}

\usepackage[hypertex,colorlinks,bookmarksnumbered=true,
linkcolor=blue,urlcolor=blue,citecolor=blue]{hyperref}

\theoremstyle{plain}
\newtheorem{statement}{Statement}

\def\SP<#1>{\langle#1\rangle}

\begin{document}
\thispagestyle{empty}
\title{Linear problems and B\"acklund transformations for the Hirota-Ohta system}
\author{V.E. Adler\thanks{L.D. Landau Institute for Theoretical Physics,
Chernogolovka, Russia.\newline E-mail: adler@itp.ac.ru},\quad
 V.V. Postnikov\thanks{Sochi Branch of Peoples' Friendship University of Russia,
Sochi, Russia.\newline E-mail: postnikovvv@rambler.ru}}

\date{26 July 2010}

\maketitle

\begin{abstract}
The auxiliary linear problems are presented for all discretization levels of
the Hirota-Ohta system. The structure of these linear problems coincides
essentially with the structure of Nonlinear Schr\"odinger hierarchy. The
squared eigenfunction constraints are found which relate Hirota-Ohta and
Kulish-Sklyanin vectorial NLS hierarchies.\medskip

Key words: auxiliary linear problem, B\"acklund transformation, discretization
%
%
\end{abstract}

\section{Introduction}\label{s:intro}

The Hirota-Ohta system \cite{Hirota_Ohta_1991} was introduced by means of
Pfaffianization applied to the Kadomtsev-Petviashvili equation, that is as a
result of a certain procedure which allows to replace multi-soliton solutions
represented by determinants with solutions represented by Pfaffians. Later on,
this procedure was applied to many other equations, see e.g.
\cite{Gilson_2002}. The differential-difference member of the Hirota-Ohta
hierarchy known as the Pfaff lattice \cite{MAdler_Horozov_Moerbeke_1999,
MAdler_Moerbeke_2002, MAdler_Shiota_Moerbeke, Kakei_1999} was introduced within
the theory of random matrix models. The whole hierarchy can be derived also
within the general approach based on Clifford algebra representations and the
boson-fermion correspondence \cite{Jimbo_Miwa, Kac_Leur}, for example, this
technique was used for the derivation of B\"acklund transformation
\cite{Leur_2004}.

In this Letter we suggest an alternative and more simple derivation of the
Hirota-Ohta hierarchy by use of Zakharov-Shabat method \cite{Zakharov_Shabat}
based on the study of auxiliary linear problems. To the best of our knowledge,
such linear problems for the continuous part of the hierarchy appeared so far
only in paper \cite{Kakei_2000}. The completely discrete Hirota-Ohta system was
found for the first time, apparently, in paper \cite{Gilson_Nimmo_Tsujimoto}.
However, it is not clear from this work, whether this discrete system belongs
to the Hirota-Ohta hierarchy or just serves as its discrete analog. In Section
\ref{s:HO}, we answer in the affirmative to this question and provide the
consistent linear problems for all discrete levels of the Hirota-Ohta system,
that is for the systems with one, two and three discrete variables.

The linear problems under consideration are rather deep reductions. For
example, in the completely discrete case, we consider a 2-component 4-point
scheme with three field variables only, while the generic scheme of such type
corresponding to the discrete Darboux-Zakharov-Manakov system
\cite{Nijhoff_Capel_1990, Bogdanov_Konopelchenko} contains eight fields (two
$2\times2$ matrices). The following empirical observation allows to recognize
these reductions:
\begin{quote}
the structure of {\em linear} problems for 3-dimensional Hirota-Ohta hierarchy
coincides with the structure of 2-dimensional hierarchy of {\em Nonlinear}
Schr\"odinger equation (NLS).
\end{quote}
The difference is that the nonlinear terms are replaced with the linear ones
with variable coefficients, the Hirota-Ohta hierarchy arising from the
compatibility conditions for these coefficients while the NLS flows commute
identically.

In Section \ref{s:NLS} we recall some standard equations from the NLS
hierarchy: the third order higher symmetry, B\"acklund transformations and the
nonlinear superposition principle. These equations are used as a hint while
choosing the form of the linear problems for the Hirota-Ohta hierarchy in
Section \ref{s:HO}. A more precise conformity with the theory of 2-dimensional
systems is pointed out in Section \ref{s:vNLS} devoted to the vectorial
generalization of NLS by Kulish and Skkyanin \cite{Kulish_Sklyanin_1981}. This
correspondence was found in \cite{Adler_2000} by one of the authors unaware of
Hirota-Ohta results at that moment.

\section{NLS hierarchy}\label{s:NLS}

The following notations are used below. The field variables $\psi,\phi$ depend
on the infinite set of independent continuous variables $t_1=x$, $t_2=y$,
$t_3=t,\dots$ and integer-valued variables
$n_1,\dots,n_i,\dots,n_j,\dots,n_k,\dots$. Subscripts $x,y,t$ denote the
partial derivatives, and subscripts $1,i,j,k$ denote the shifts on the discrete
variables: $\psi_i=\psi(\dots,n_i+1,\dots)$, minus sign denoting the backward
shift: $\psi_{-i}=\psi(\dots,n_i-1,\dots)$.

The Nonlinear Schr\"odinger equation is of the form
\begin{equation}\label{NLS.y}
 \psi_y=\psi_{xx}+2\psi^2\phi,\quad -\phi_y=\phi_{xx}+2\phi^2\psi.
\end{equation}
Its simplest higher symmetry is of third order:
\begin{equation}\label{NLS.t}
 \psi_t=\psi_{xxx}+6\phi\psi\psi_x,\quad \phi_t=\phi_{xxx}+6\psi\phi\phi_x.
\end{equation}
The discrete variable $n_1$ will play the distinguished role in what follows.
The corresponding shift defines B\"acklund-Schlesinger transformation as the
explicit mapping
\begin{equation}\label{NLS.B1}
 \psi_1=\psi_{xx}-\psi^2_x/\psi+\psi^2\phi,\quad \phi_1=1/\psi
\end{equation}
which acts on the solutions of systems (\ref{NLS.y}), (\ref{NLS.t}). The
iterations of this mapping are governed, under the change $\psi=e^q$,
$\phi=e^{-q_{-1}}$, by the Toda lattice
\begin{equation}\label{NLS.qxx}
 q_{xx}=e^{q_1-q}-e^{q-q_{-1}}.
\end{equation}
All other discrete variables correspond to B\"acklund-Darboux transformations;
$x$-parts of these transformations are of the form
\begin{equation}\label{NLS.Bi}
 \psi_x=\psi_i+\alpha^{(i)}\psi+\psi^2\phi_i,\quad
 -\phi_{i,x}=\phi+\alpha^{(i)}\phi_i+\phi^2_i\psi,
\end{equation}
where $\alpha^{(i)}$ is an arbitrary parameter, associated with $i$-th discrete
coordinate direction and depending on $n_i$ only (that is,
$\alpha^{(i)}_j=\alpha^{(i)}$ at $j\ne i$). Second equation (\ref{NLS.Bi})
defines $\phi_i$ as a solution of Riccati equation, then $\psi_i$ is explicitly
found from the first equation. A generic B\"acklund transformation for the NLS
equation is decomposed as a sequence of elementary transformations of the form
(\ref{NLS.B1}), (\ref{NLS.Bi}) and their inverses.

The completely discrete part of the NLS hierarchy is described by 5-point
equations of discrete Toda type
\begin{equation}\label{NLS.1i}
 e^{q_{1,-i}-q}-e^{q-q_{-1,i}}+e^{q_i-q}-e^{q-q_{-i}}+\alpha^{(i)}-\alpha^{(i)}_{-i}=0
\end{equation}
and 2-component quad-equations
\begin{equation}\label{NLS.ij}
 \psi_j=\psi_i+\frac{(\alpha^{(i)}-\alpha^{(j)})\psi}{1-\psi\phi_{ij}},\quad
 \phi_j=\phi_i+\frac{(\alpha^{(j)}-\alpha^{(i)})\phi_{ij}}{1-\psi\phi_{ij}}.
\end{equation}
Equations (\ref{NLS.1i}) define the permutability property of the
transformations (\ref{NLS.B1}) and (\ref{NLS.Bi}). This means that equation
(\ref{NLS.1i}) is consistent with the differentiation
\[
 q_x=e^{q_i-q}+e^{q-q_{-1,i}}+\alpha^{(i)}
    =e^{q_{1,-i}-q}+e^{q-q_{-i}}+\alpha^{(i)}_{-i},
\]
that is the derivative with respect to $x$ of the left-hand side of
(\ref{NLS.1i}) vanishes in virtue of (\ref{NLS.1i}) itself. Moreover, the
variables $q(n_1,n_i)$ at fixed $n_i$ satisfy the Toda lattice (\ref{NLS.qxx}),
and variables $\psi=e^{q(n_1,n_i)}$, $\phi=e^{-q(n_1-1,n_i)}$ at fixed $n_1$
satisfy the lattice (\ref{NLS.Bi}).

\begin{figure}[t]
\centerline{\includegraphics[width=0.8\textwidth]{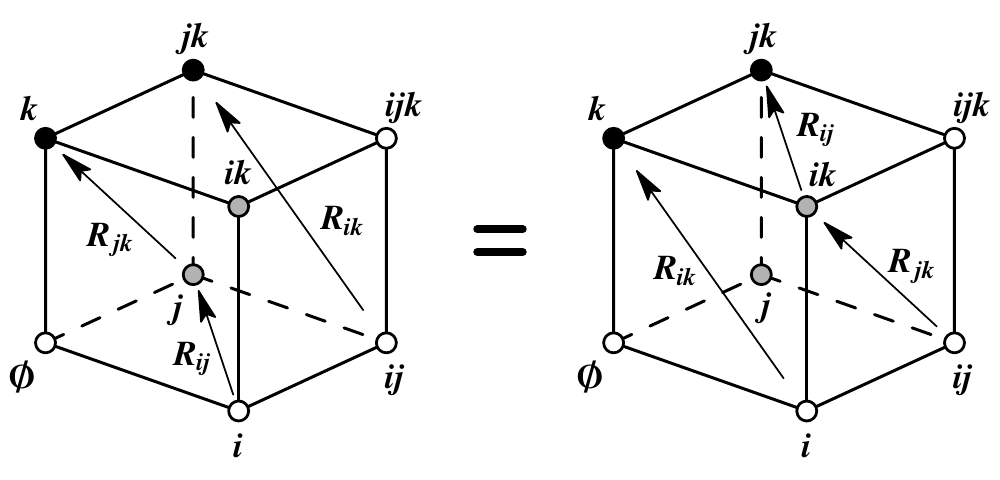}}
\caption{$R_{jk}R_{ik}R_{ij}=R_{ij}R_{ik}R_{jk}$}\label{fig:RRR}
\end{figure}

Analogously, equations (\ref{NLS.ij}) define the permutability property of two
transformations of the form (\ref{NLS.Bi}). Let $\Psi=(\psi,\phi)$, then the
following property holds: if $\Psi$ and $\Psi_i$ are related by the
transformation (\ref{NLS.Bi}) with parameter $\alpha^{(i)}$, and $\Psi_i$ and
$\Psi_{ij}$ are related by the transformation with parameter $\alpha^{(j)}$
then the variable $\Psi_j$ defined accordingly to (\ref{NLS.ij}) is as well
related to $\Psi$ and $\Psi_{ij}$ by the transformations of the same type, but
with the interchanged parameters $\alpha^{(j)}$ and $\alpha^{(i)}$.

The mappings $R_{ij}:(\Psi,\Psi_i,\Psi_{ij})\mapsto\Psi_j$ satisfy an important
3D-consistency property expressed by the Yang-Baxter identity (see fig.
\ref{fig:RRR})
\begin{equation}\label{YB}
 R_{jk}R_{ik}R_{ij}=R_{ij}R_{ik}R_{jk}.
\end{equation}
An essential distinction with the scalar quad-equations
\cite{Adler_Bobenko_Suris_2003} is that equations (\ref{NLS.ij}) cannot be
solved with respect to variables $\Psi$ or $\Psi_{ij}$. Hence the formulation
of 3D-consistency is possible here only under a choice of the initial data
along some sequence of the vertices of the cube, for example $\Psi$, $\Psi_1$,
$\Psi_{12}$, $\Psi_{123}$.

\section{Hirota-Ohta hierarchy}\label{s:HO}

We keep the previous notations for the derivatives and shifts with respect to
the independent variables. As for the dependent variables, now $\psi$ and
$\phi$ play the role of wave-functions, while the field variables will be
denoted $u,v$, $w,w^{(i)},w^{(ij)},q$ (coefficients of linear problems) and
$f,g,h$ ($\tau$-functions). Superscripts in $w^{(i)}$, $w^{(ij)}$ indicate that
these variables are associated with the corresponding direction or plane in the
lattice (but, in contrast to parameters $\alpha^{(i)}$ from the previous
section, these depend on the whole set of independent variables).

We accept the following linearization of the NLS equation (\ref{NLS.y}) as a
starting point:
\begin{equation}\label{HO.y}
 \psi_y=\psi_{xx}+w\psi+2u\phi,\quad -\phi_y=\phi_{xx}+2v\psi+w\phi.
\end{equation}
Now, let us search for a commuting flow of the form
\[
 \psi_t=\psi_{xxx}+a\psi_x+b\psi+c\phi,\quad \phi_t=\phi_{xxx}+A\phi_x+B\psi+C\phi
\]
with undetermined coefficients (that is, the cubic terms in (\ref{NLS.t}) are
replaced with the linear combination of the factors). An easy computation
specifies the coefficients:
\begin{equation}\label{HO.t}
\begin{aligned}[b]
 \psi_t&=\psi_{xxx}+\frac32w\psi_x+\frac34(w_x+q)\psi+3u_x\phi,\\
 \phi_t&=\phi_{xxx}+\frac32w\phi_x+\frac34(w_x-q)\phi+3v_x\psi
\end{aligned}
\end{equation}
and yields the compatibility conditions of equations (\ref{HO.y}) and
(\ref{HO.t}) in the form of the following system:
\begin{equation}\label{HO}
\begin{aligned}[b]
 -2u_t&=u_{xxx}-3u_{xy}+3wu_x-3qu,\\
 -2v_t&=v_{xxx}+3v_{xy}+3wv_x+3qv,\\
  4w_t&=w_{xxx}-24(uv)_x+6ww_x+3q_y,\quad q_x=w_y.
\end{aligned}
\end{equation}
This is nothing but the Hirota-Ohta system \cite{Hirota_Ohta_1991}. It is clear
that its higher symmetries can be derived quite analogously by use of the
linear problems of higher order with respect to $\partial_x$; however, we will
skip this.

The change of variables
\begin{equation}\label{fgh}
 u=\frac{f}{h},\quad v=\frac{g}{h},\quad w=2(\log h)_{xx},\quad q=2(\log h)_{xy}
\end{equation}
brings the system (\ref{HO}) to the bilinear form
\begin{equation}\label{HO.2}
\begin{aligned}[b]
 & (2D_t-3D_xD_y+D^3_x)f\cdot h=0,\\
 & (2D_t+3D_xD_y+D^3_x)g\cdot h=0,\\
 & (4D_xD_t-3D^2_y-D^4_x)h\cdot h+24fg=0
\end{aligned}
\end{equation}
where $D_z a\cdot b=a_zb-ab_z$ denotes the Hirota operator. The variables
$f,g,h$ are especially convenient for rewriting of the discrete dynamics.

It is natural to search for an analog of the transformation (\ref{NLS.B1}) in
the form
\[
 \psi_1=a\psi_{xx}+b\psi_x+c\psi+d\phi,\qquad \phi_1=p\psi.
\]
The answer is not unique, because of the gauge transformations
\[
 \psi=\alpha\mu(y)\tilde\psi,\quad
 \phi=\frac{\beta}{\mu(y)}\tilde\phi,\quad
 \tilde u=\frac{\beta}{\alpha\mu^2}u,\quad
 \tilde v=\frac{\alpha\mu^2}{\beta}v,\quad
 \tilde w=w-\frac{\mu_y}{\mu}
\]
which leave invariant the form of basic linear problem (\ref{HO.y}). It is
possible to fix some coefficients by use of these transformations and to choose
the mapping in the following form:
\begin{equation}\label{HO.1}
 \psi_1=\psi_{xx}-\frac{u_x}{u}\psi_x+\Bigl(w+\frac{u_{xx}-u_y}{2u}\Bigr)\psi+u\phi,
 \quad \phi_1=-\frac{1}{u}\psi.
\end{equation}
Its action on the coefficients of the linear problems is described as follows.

\begin{statement}
System (\ref{HO}) admits the explicit auto-transformation
\begin{equation}\label{HO.B1}
\begin{gathered}[b]
 u_1= u^2v+\frac12(u_xw_x-uw_y)+\frac{w}{u}(uu_{xx}-u^2_x)
      +\frac1{4u}\bigl(uu_{yy}-u^2_y\\
      -2uu_{xxy}+2u_xu_{xy}+uu_{xxxx}-2u_xu_{xxx}+u^2_{xx}\bigr),\\[1ex]
 v_1=1/u,\qquad w_1=w+2(\log u)_{xx},\qquad q_1=q+2(\log u)_{xy}.
\end{gathered}
\end{equation}
\end{statement}

The substitutions (\ref{fgh}) reduce the last three equations in (\ref{HO.B1})
to $h_1=f$, $g_1=h$, that is the iterations of this mapping generate the
sequence
\[
 \dots\quad f=h(n_1+1),\quad h=h(n_1),\quad g=h(n_1-1)\quad \dots
\]
Moreover, the first equation (\ref{HO.B1}) is equivalent to the relation
\[
 4\Bigl(\frac{A}{h_1h}\Bigr)_x=\frac{B_1}{h_1^2}-\frac{B}{h^2},
\]
where
\[
 A=(3D_xD_y-D^3_x)h_1\cdot h,\quad B=(3D^2_y+D^4_x)h\cdot h-24h_1h_{-1}.
\]
This relation is not bilinear, however it turns into identity in virtue of the
system (\ref{HO.2}) which takes the form (cf
\cite{MAdler_Horozov_Moerbeke_1999, MAdler_Moerbeke_2002,
MAdler_Shiota_Moerbeke})
\begin{equation}\label{HO.h}
\begin{aligned}[b]
 & (2D_t-3D_xD_y+D^3_x)h_1\cdot h=0,\\
 & (4D_xD_t-3D^2_y-D^4_x)h\cdot h+24h_1h_{-1}=0.
\end{aligned}
\end{equation}
Indeed, in virtue of these equations
\[
 \frac{A}{h_1h}=2r_1-2r,\quad \frac{B}{h^2}=8r_x,\quad r=\frac{h_t}{h}.
\]

Darboux-B\"acklund transformation is derived from the linear problem
\begin{equation}\label{HO.i}
 \psi_x=\psi_i+w^{(i)}\psi+u\phi_i,\quad -\phi_{i,x}=\phi+v_i\psi+w^{(i)}\phi_i,
\end{equation}
and an analog of the nonlinear superposition principle (\ref{NLS.ij}) is
described by the linear problem
\begin{equation}\label{HO.ij}
 \psi_j=\psi_i+w^{(ij)}(\psi+u\phi_{ij}),\quad
 \phi_j=\phi_i-w^{(ij)}(v_{ij}\psi+\phi_{ij}),\quad i\le j.
\end{equation}
The compatibility conditions for these equations define all discrete levels of
the Hirota-Ohta system. We write them in the bilinear form at once, since the
equations for the coefficients are rather bulky. To do this we use the
following relations, in addition to the substitutions (\ref{fgh}):
\begin{equation}\label{ww}
 w^{(i)}=\frac{h_{i,x}}{h_i}-\frac{h_x}{h},\quad w^{(ij)}=\frac{hh_{ij}}{h_ih_j}.
\end{equation}

\begin{statement}
The compatibility conditions for equations (\ref{HO.y}) and (\ref{HO.i}) yield
the system
\begin{alignat}{2}
\nonumber
 &fh_{i,y}-f_yh_i+fh_{i,xx}-2f_xh_{i,x}+f_{xx}h_i-2f_ih&&=0,\\
\nonumber
 &hg_{i,y}-h_yg_i+hg_{i,xx}-2h_xg_{i,x}+h_{xx}g_i-2h_ig&&=0,\\
\label{HO.ixy}
 &hh_{i,y}-h_yh_i-hh_{i,xx}+2h_xh_{i,x}-h_{xx}h_i+2fg_i&&=0;
\end{alignat}
the compatibility conditions of (\ref{HO.ij}) and two equations of the form
(\ref{HO.i}), corresponding to $i$ and $j$, bring to the system
\begin{alignat}{2}
\nonumber
 &fh_{ij,x}-f_xh_{ij}&&=f_ih_j-f_jh_i,\\
\nonumber
 &hg_{ij,x}-h_xg_{ij}&&=h_ig_j-h_jg_i,\\
\label{HO.ijx}
 &h_ih_{j,x}-h_{i,x}h_j&&=fg_{ij}-hh_{ij},\quad i<j;
\end{alignat}
the compatibility conditions for three linear problems of the form
(\ref{HO.ij}), corresponding to $i,j$ and $k$, yield the system
\begin{alignat}{2}
\nonumber
 &fh_{ijk}-f_ih_{jk}+f_jh_{ik}-f_kh_{ij}&&=0,\\
\nonumber
 &hg_{ijk}-h_ig_{jk}+h_jg_{ik}-h_kg_{ij}&&=0,\\
\label{HO.ijk}
 &fg_{ijk}-h_ih_{jk}+h_jh_{ik}-h_kh_{ij}&&=0,\quad i<j<k.
\end{alignat}
\end{statement}

Equations (\ref{HO.ijk}) are nothing but the discrete version of the
Hirota-Ohta system found in \cite{Gilson_Nimmo_Tsujimoto}. Let us explain its
derivation in more details. Formally one may consider the linear problem of
form (\ref{HO.ij}) as a very special case of 4-point scheme for 2-component
function $\Psi=(\psi,\phi)$. As in the case of the NLS equation, $\phi$ and
$\psi_{ij}$ are not involved in the equations, and this dictates the following
choice of the initial data needed for the computation of compatibility
conditions: $\Psi,\Psi_i,\Psi_{ij},\Psi_{ijk}$. For the sake of definiteness we
will assume that $i<j<k$. Then the compatibility condition is expressed by
Yang-Baxter equation (\ref{YB}) where $R_{ij}$ is the mapping
$(\Psi,\Psi_i,\Psi_{ij})\mapsto\Psi_j$ defined by equations (\ref{HO.ij}). The
computations use the following equations:
\begin{alignat*}{3}
 R_{ij}:&~&&
 \begin{array}{ll}
  \psi_j=\psi_i+w^{(ij)}(\psi+u\phi_{ij}),\\
  \phi_j=\phi_i-w^{(ij)}(v_{ij}\psi+\phi_{ij}),
 \end{array} &~&
 \begin{array}{ll}
  \psi_{jk}=\psi_{ik}+w^{(ij)}_k(\psi_k+u_k\phi_{ijk}),\\
  \phi_{jk}=\phi_{ik}-w^{(ij)}_k(v_{ijk}\psi_k+\phi_{ijk}),
 \end{array}\\[1ex]
 R_{ik}:&&&
 \begin{array}{ll}
  \psi_{jk}=\psi_{ij}+w^{(ik)}_j(\psi_j+u_j\phi_{ijk}),\\
  \phi_{jk}=\phi_{ij}-w^{(ik)}_j(v_{ijk}\psi_j+\phi_{ijk}),
 \end{array}&&
 \begin{array}{ll}
  \psi_k=\psi_i+w^{(ik)}(\psi+u\phi_{ik}),\\
  \phi_k=\phi_i-w^{(ik)}(v_{ik}\psi+\phi_{ik}),
 \end{array}
 \\[1ex]
 R_{jk}:&&&
 \begin{array}{ll}
  \psi_k=\psi_j+w^{(jk)}(\psi+u\phi_{jk}),\\
  \phi_k=\phi_j-w^{(jk)}(v_{jk}\psi+\phi_{jk}),
 \end{array} &&
 \begin{array}{ll}
  \psi_{ik}=\psi_{ij}+w^{(jk)}_i(\psi_i+u_i\phi_{ijk}),\\
  \phi_{ik}=\phi_{ij}-w^{(jk)}_i(v_{ijk}\psi_i+\phi_{ijk}).
 \end{array}
\end{alignat*}
Left hand side of Yang-Baxter equation (\ref{YB}) corresponds to the
consecutive computations according to the left column downwards, while the
right hand side corresponds to the right column computed upwards. The resulting
nonlinear relations for the coefficients contain, in particular, the
conservation laws which imply the parametrization of $w^{(ij)}$ given in
(\ref{ww}). The rest relations then turn out to be equivalent to system
(\ref{HO.ijk}).

\section{Kulish-Sklyanin vector hierarchy}\label{s:vNLS}

A wide and well-known class of reductions from 3-dimensional equations to
vectorial 2-dimensional ones consists of so-called squared eigenfunction
constraints. For instance, the Manakov system \cite{Manakov,Fordy_Kulish} and
its third order symmetry
\begin{gather*}
   \psi_y= \psi_{xx} +2\SP<\psi,\phi>\psi, \quad
  -\phi_y= \phi_{xx} +2\SP<\psi,\phi>\phi, \\
   \psi_t= \psi_{xxx} +3\SP<\psi,\phi>\psi_x +3\SP<\psi_x,\phi>\psi,\quad
   \phi_t= \phi_{xxx} +3\SP<\psi,\phi>\phi_x +3\SP<\psi,\phi_x>\phi
\end{gather*}
define such a reduction for the Kadomtsev-Petviashvili equation
\[
  4u_t=u_{xxx}-6uu_x+3q_y, \quad q_x=u_y
\]
with respect to the quantities $u=-2\SP<\psi,\phi>$,
$q=2\SP<\psi,\phi_x>-2\SP<\psi_x,\phi>$ \cite{Konopelchenko_Sidorenko_Strampp}.

Now we demonstrate that the Hirota-Ohta hierarchy (for all discrete levels)
arises analogously from another vectorial version of the NLS equation
introduced by Kulish and Sklyanin \cite{Kulish_Sklyanin_1981}:
\begin{equation}\label{KS}
\begin{aligned}[b]
  \psi_y&=\psi_{xx}+4\SP<\psi,\phi>\psi-2\SP<\psi,\psi>\phi,\\
 -\phi_y&=\phi_{xx}+4\SP<\psi,\phi>\phi-2\SP<\phi,\phi>\psi.
\end{aligned}
\end{equation}
Let us write the corresponding generalizations for the basic equations in
Section \ref{s:NLS}. The third order symmetry reads
\begin{equation}\label{KS3}
\begin{aligned}[b]
 \psi_t&=\psi_{xxx}+6\SP<\psi,\phi>\psi_x+6\SP<\psi_x,\phi>\psi-6\SP<\psi,\psi_x>\phi,\\
 \phi_t&=\phi_{xxx}+6\SP<\psi,\phi>\phi_x+6\SP<\psi,\phi_x>\phi-6\SP<\phi,\phi_x>\psi.
\end{aligned}
\end{equation}
The analog of B\"acklund-Schlesinger transformation (\ref{NLS.B1}) is of the form
\cite{Svinolupov_Yamilov_1994}
\begin{equation}\label{KS.B1}
\begin{aligned}[b]
 \psi_1&=\psi_{xx}-2\frac{\SP<\psi,\psi_x>}{\SP<\psi,\psi>}\psi_x
 +\frac{\SP<\psi_x,\psi_x>}{\SP<\psi,\psi>}\psi+2\SP<\psi,\phi>\psi-\SP<\psi,\psi>\phi,\\
 \phi_1&=\frac1{\SP<\psi,\psi>}\psi,
\end{aligned}
\end{equation}
the analog of B\"acklund-Darboux transformation (\ref{NLS.Bi}) reads
\begin{equation}\label{KS.Bi}
\begin{aligned}[b]
 \psi_x&=\psi_i+\alpha^{(i)}\psi+2\SP<\psi,\phi_i>\psi-\SP<\psi,\psi>\phi_i,\\
 -\phi_{i,x}&=\phi+\alpha^{(i)}\phi_i+2\SP<\psi,\phi_i>\phi_i-\SP<\phi_i,\phi_i>\psi
\end{aligned}
\end{equation}
and, finally, the analog of nonlinear superposition principle (\ref{NLS.ij}) is
\cite{Adler_1994}
\begin{equation}\label{KS.ij}
\begin{aligned}[b]
 \psi_j=\psi_i+\frac{(\alpha^{(i)}-\alpha^{(j)})(\psi-\SP<\psi,\psi>\phi_{ij})}
                    {1-2\SP<\psi,\phi_{ij}>+\SP<\psi,\psi>\SP<\phi_{ij},\phi_{ij}>},\\
 \phi_j=\phi_i-\frac{(\alpha^{(i)}-\alpha^{(j)})(\phi_{ij}-\SP<\phi_{ij},\phi_{ij}>\psi)}
                    {1-2\SP<\psi,\phi_{ij}>+\SP<\psi,\psi>\SP<\phi_{ij},\phi_{ij}>}.
\end{aligned}
\end{equation}
The comparison with the linear problems from the previous section proves
immediately that the Kulish-Sklyanin hierarchy is their self-consistent
reduction. More precisely, the following statement holds true.

\begin{statement}
Equations (\ref{KS})--(\ref{KS.ij}) are consistent and, in virtue of these
equations, the quantities
\begin{gather*}
 u=-\SP<\psi,\psi>,\quad
 v=-\SP<\phi,\phi>,\quad
 w=4\SP<\psi,\phi>,\quad
 q=4\SP<\psi_x,\phi>-4\SP<\psi,\phi_x>,\\
 w^{(i)}=\alpha^{(i)}+2\SP<\psi,\phi_i>,\quad
 w^{(ij)}=\frac{\alpha^{(i)}-\alpha^{(j)}}
  {1-2\SP<\psi,\phi_{ij}>+\SP<\psi,\psi>\SP<\phi_{ij},\phi_{ij}>}
\end{gather*}
satisfy the equations of the Hirota-Ohta hierarchy.
\end{statement}

\section{Concluding remarks}

The main result of this Letter is an uniform derivation of continuous and
discrete equations of the Hirota-Ohta hierarchy from the compatibility
conditions of auxiliary linear problems. The structure of these problems
patterns after the structure of 2-dimensional NLS hierarchy. We hope that this
observation may be useful for other 3-dimensional equations as well.

It is worth noticing also the relation of above discrete linear problems with
self-adjoint 5- and 7-point schemes, on square and triangular lattices
respectively. The linear problems of this type attracted much attention
recently. In particular, it is known that they appear from the scalar 4-point
schemes on square and rhombic lattices as a result of restriction on even/odd
sublattices \cite{Doliwa_Nieszporski_Santini_2007}. In our case, such linear
problems appear from 2-component linear problems presented in Section
\ref{s:HO} as a result of elimination of one component. For instance, the
elimination of $\phi$ from (\ref{HO.1}) and (\ref{HO.i}) yields the following
5-point equation (compare with discrete Toda lattice (\ref{NLS.1i})):
\[
 \frac1{u}\psi_i+\frac1{u_{-i}}\psi_{-i}
 -\frac1{u}\psi_{1,-i}-\frac1{u_{-1,i}}\psi_{-1,i}+
 \frac1{u}\Bigl(w^{(i)}_{1,-i}+w^{(i)}-\frac{u_x}{u}\Bigr)\psi=0.
\]
Analogously, eliminating $\phi$ from equations (\ref{HO.ij}) results in the
7-point linear problem in the plane $(ij)$. We mention that some examples of
continuous Toda-like dynamics associated with 5- and 7-point linear problems
were studied in the papers \cite{Santini_Nieszporski_Doliwa_2004,
Gilson_Nimmo_2005, Santini_Doliwa_Nieszporski_2008}. It remains unclear,
whether these examples coincide with the flow (\ref{HO.i}) of the Hirota-Ohta
hierarchy; more likely, they may interpreted as its negative flows.

\paragraph{Acknowledgements.} The research of V.A. was supported by RFBR grants
08-01-00453, 09-01-92431-KE and NSh-6501.2010.2.

\def\doi#1#2{\href{http://dx.doi.org/#1}{#2}}

\end{document}